\documentstyle[psfig,twocolumn,aps]{revtex}
\begin{document}

\draft
\title{Metals with Small Electron Mean-Free Path: Saturation
{\sl versus} Escalation of Resistivity}

\author{ Philip B. Allen}
\address{Department of Physics and Astronomy, State University of New York,
Stony Brook, New York 11794-3800}

\date{\today}

\maketitle

\begin{abstract}
\begin{center} Abstract
\end{center}
Resistivity of metals is commonly observed either to `escalate'
beyond the Ioffe-Regel limit (mean free path $\ell$ equal
to lattice constant $a$) or to `saturate' at this point.
It is argued that neither behavior is well-understood, and
that `escalation' is not necessarily more mysterious than `saturation.'

\end{abstract}



\begin{figure}[t]
\centerline{\psfig{figure=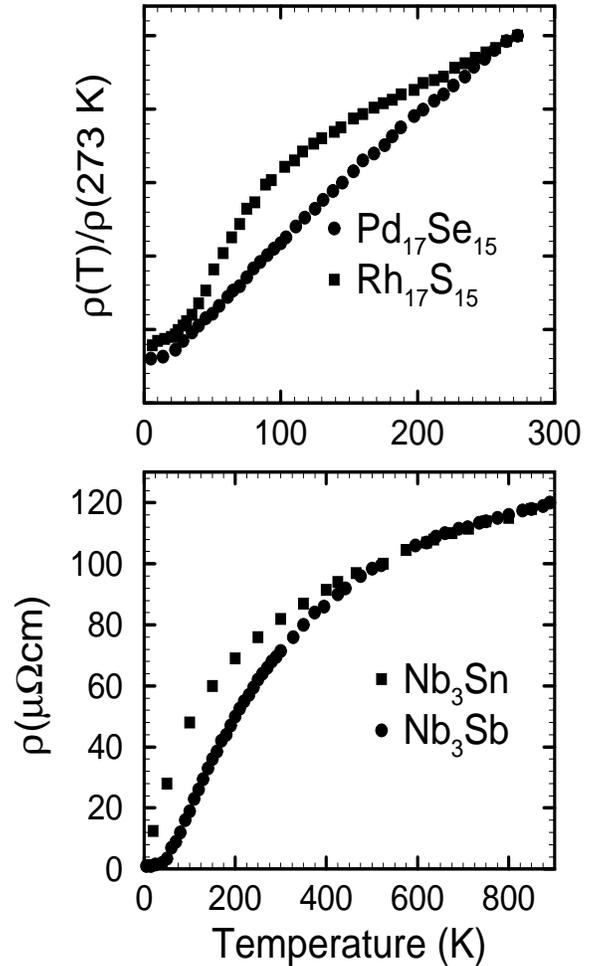,height=5.0in,width=3.4in,angle=0}}
\caption{Upper panel: resistivity from Ref.
\protect\cite{Fisk1} normalized to the value at 273 K.  
The `bulge' in Rh$_{17}$S$_{15}$ was originally
correlated with the higher superconducting
transition.  Lower panel: absolute resistivity from Ref.
\protect\cite{Fisk2}
of a good superconductor (Nb$_3$Sn, $T_c$=18 K) saturates
at the same final resistivity value as a poor superconductor (Nb$_3$Sb).
This shows that `saturation' correlates with
magnitude of resistivity rather than $T_c$ 
or strength of electron-phonon coupling.}
\label{fig:fisk}
\end{figure}

During the period 1954-1986, `high temperature' superconductivity
occured primarily in Nb$_3$Sn and related materials with the
A15 crystal structure.  Although peculiar resistivity $\rho(T)$
had been noticed \cite{Woodward} and modeled (corrected for
non-zero $k_B T/\epsilon_F$) \cite{Cohen}, it was not until the
Fisk-Lawson paper \cite{Fisk1} that the general nature of the
peculiarity was known.  At first the peculiarity seemed to correlate
with high $T_c$ and therefore large electron-phonon coupling, but
the Fisk-Webb paper \cite{Fisk2} showed that even a low $T_c$
material had `saturating' resistivity when the resistivity
became large enough.  Data from these two classic papers
is shown in Fig. \ref{fig:fisk}.
Fisk and Webb observed that the mean free path $\ell$
was becoming nearly as small as the lattice constant $a$.  This
violation of the condition $\ell \gg a$ (required
for validity of the quasiparticle picture and the Bloch-Boltzmann
resistivity theory) seemed from the available data 
to lead quite generally to
`saturation' of resistivity.  
Unfortunately, correcting theory for non-zero $a/\ell$ (with
values of order 1) is more important, and
much more difficult than correcting
for non-zero $k_B T/\epsilon_F$ (which is usually small.)

To strengthen the case,
`saturation' could be provoked not only
by the thermal disorder of high $T$ lattice vibrations, but also
by static disorder, such as radiation damage \cite{Gurvitch}.
It could also be modelled by a `shunt-resistor model'
\cite{Wiesmann} $1/\rho(T)=1/\rho_{\rm Boltz}+1/\rho_{\rm sat}$.
If we take the saturation value $\rho_{\rm sat}$ to be the
same as the Boltzmann value $\rho_{\rm Boltz}$ extrapolated
to the Ioffe-Regel point $\ell=a$, then the `shunt-resistor'
formula can be written $\rho(T)=\rho_{\rm Boltz}(T)/(1+a/\ell(T))$.
When $a/\ell(T)$ is large and the second term dominates the
denominator, then $\ell(T)$ cancels from the formula and $\rho$
becomes independent of $T$.

In spite of the explicit warning \cite{Fisk2} that no theory
of `saturation' existed, it was often assumed that the
case was closed.  The reasoning \cite{Gurvitch2} 
was that $\ell$ cannot sensibly
become smaller than $a$, so one should just use the Boltzmann
result with $\ell$ replaced by $a$.  The fault with this argument
\cite{Allen}, that it denies the possibility of Anderson localization
under static disorder, was easy to forget.  A careful
theoretical discussion has to mention that if $\ell$ is not
longer than $a$, then the concept of a mean free path has to be
abandoned.  One has no right to take a formula which contains
a factor $1/\ell$ and replace it with $1/a$, because the theoretical
framework which provided the formula has already been destroyed.

\par
\par

\begin{figure}[t]
\centerline{\psfig{figure=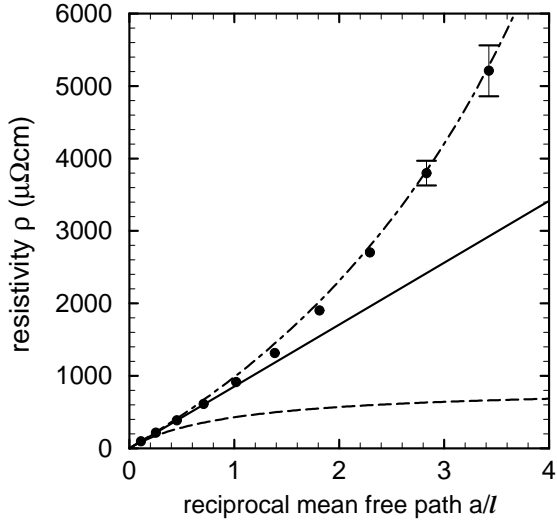,height=3.0in,width=3.4in,angle=0}}
\caption{Resistivity of a half-filled
$s$-band on a cubic lattice with random on-site
potential fluctuations, plotted {\sl versus}
$a/\ell$.  Here $\ell$ is not a physical
mean free path, but instead the value given by Boltzmann
theory, extended beyond is range of validity.
The points were calculated in ref. 
\protect\cite{Nikolic}.
The solid line is the Bloch-Boltzmann prediction in Born approximation.  
The dashed line is the expected saturated form, $\rho_{\rm Boltz}/(1+a/\ell)$,
and the dot-dashed line is a naive scaling formula 
$\rho_{\rm Boltz}/(1-\ell_c/\ell)$, with $\ell_c=a/7.71$, the observed
point of the Anderson transition.}
\label{fig:nikolic}
\end{figure}

In recent years it became possible to calculate resistivity
for highly disordered metals (with non-interacting electrons)
by the Kubo formula or an equivalent
Landauer formulation.  The results \cite{Nikolic} for the most familiar model
are shown in Fig. \ref{fig:nikolic}.  Rather than saturating,
the resistivity escalates.  In the center of the band formed
by a single $s$ orbital on a simple cubic lattice, Boltzmann
theory (in Born approximation, at $T=0$) gives \cite{Nikolic}
$\rho_{\rm Boltz}=(\hbar a/e^2)(6.94a/\ell)$, and 
$a/\ell=0.0283(W/t)^2$, where $t$ is the hopping parameter and
$\pm W/2$ is the interval of the random on-site disorder potential.
Numerically converged results for large disorder are plotted
versus $a/\ell$, with $a$ set arbitrarily
at $3\AA$.  Of course, $\ell$ is meaningless over
most of the range shown, and the horizontal axis is just
an alternate parameterization of $(W/t)^2$.  There are three
things to notice: (1) Boltzmann theory continues to give 
a good answer even at $a/\ell=1$ where it is invalid; (2)
there is a very large interval between the Ioffe-Regel point
$a/\ell=1$ and the point where states become all localized,
$(a/\ell)_c=7.71$ \cite{Slevin}; (3) rather than saturating,
the resistivity escalates toward the scaling region where it
diverges as $((a/\ell)_c -a/\ell)^{-\nu}$ with critical
exponent $\nu\approx 1.57$.

\par
\par

\begin{figure}[t]
\centerline{\psfig{figure=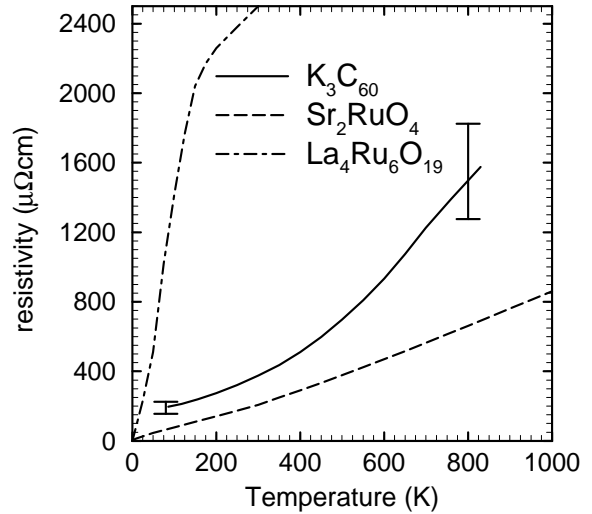,height=3.0in,width=3.4in,angle=0}}
\caption{Resistivity {\it versus} temperature for various metals.
Data are from Ref. \protect\cite{Hou} (K$_3$C$_{60}$);
Ref. \protect\cite{Pavuna} (Sr$_2$RuO$_4$);
Ref. \protect\cite{Khalifah} (La$_4$Ru$_6$O$_{19}$).  The error bars
represent the error estimated in the geometric normalization for
K$_3$C$_{60}$.}
\label{fig:various}
\end{figure}

One of the first indications that `saturation' was not ubiquitous
was the observation by Gurvitch and Fiory \cite{Gurvitch3} that 
resistivity of high $T_c$ superconductors seems to fail to saturate.
The failure of `saturation', which I will call `escalation',
is probably not rare after all.  Emery and Kivelson \cite{Emery}
call such metals `bad metals.'
Fig. \ref{fig:various} shows resistivities
of several metals.  Alkali-doped C$_{60}$ saturates (if at all) \cite{Hou}
only for very short $\ell$) (less than $1\AA$.)  Among oxide metals,
saturation seems rare for simpler compounds like Sr$_2$RuO$_4$
\cite{Pavuna,Tyler}, and may occur at very large resistivity values
for certain complex oxides like La$_4$Ru$_6$O$_{19}$ \cite{Khalifah}.
Quite likely the mean free path in this last case is also very small.

\par

There are exotic theories which attribute \cite{Kiv} `bad metal' behavior
to quasi-1d electron conduction on fluctuating stripes.  One can naively 
model this by meandering stripes containing metallic electrons whose
resistivity is not high. 
Dilution of the metallic stripes in an intervening non-metallic
phase causes an apparent high resistivity and short mean-free path.
Arnason {\it et al.} \cite{Arnason} reported 
an interesting experiment where a static realization
of this geometry was intentionally created.

Several recent theories \cite{Millis,Merino,Gunnarsson1,Gunnarsson2}
attempt to find or explain saturation behavior.
Millis {\sl et al.} \cite{Millis} find a tendency in the direction of saturation
by applying the `dynamical mean field theory' approximation to
a model with electron-phonon interactions and disorder, but
no Coulomb scattering.  Merino and McKenzie \cite{Merino}
find a similar effect in a model with on-site (Hubbard)
Coulomb repulsion and no phonons.  Both theories have only a single
band.  Gunnarsson and Han \cite{Gunnarsson1} use a 3-fold degenerate
band to model doped C$_{60}$, and Calandra and Gunnarsson
\cite{Gunnarsson2} use a 5-fold degenerate band to model
Nb$_3$X.  These interesting quantum-Monte-Carlo (QMC) calculations
find no saturation in the C$_{60}$ model, and saturation
in the Nb$_3$X model.  The difference apparently is in the
form of electron-phonon coupling used (Coulomb interactions
were omitted).  When phonons were coupled to on-site potential
as in Fig. \ref{fig:nikolic}, saturation was not found,
whereas when coupled to the hopping matrix element, vibrations
were found to cause saturating resistivity.  In this latter
case, theoretical QMC results could be fit by Kubo theory
calculated under the assumption that vibrations could be treated
adiabatically, {\it i. e.}
modelled as static disorder.  This approximate analysis 
agrees with qualitative arguments I made with Chakraborty long ago
\cite{Chakraborty}, emphasizing the importance of allowing
a multi-orbital band structure.

The conclusions to draw from this are: (1) there still is
no theory of saturation, although computational results are
giving a first outline; (2) it seems no harder to account
for escalation than for saturation.  Therefore it is
perhaps an unproven guess that metals with escalating
resistivity are more exotic than metals
which show saturation.  However, an exotic origin of
escalation has been nicely demonstrated in the case of
high $T_c$ cuprates.  Resistivity versus doping and
temperature was re-examined by Ando {\it et al.} \cite {Ando}.
They make a very convincing case that the mechanism of
transport is intimately related to antiferromagnetic 
correlations over a wide doping range.  A candidate
picture by which this can happen is the model of
self-organizing stripe inhomogeneities \cite{Kiv}.

\acknowledgements
I thank Z. Fisk and B. Nikoli\'c for their many contributions
to this work, and S. A. Kivelson for very helpful comments on
the manuscript.  This work was supported by NSF grant no. DMR-0089492.



\begin{references}

\bibitem{Woodward}	D. W. Woodward and G. D. Cody, 
			Phys. Rev. {\bf 136} (1964) A166.

\bibitem{Cohen}		R. W. Cohen, G. D. Cody, and J. J. Halloran,
			Phys. Rev. Lett. {\bf 19} (1967) 840.

\bibitem{Fisk1}       Z. Fisk and A. C. Lawson,
			Solid State Commun. {\bf 13} (1973) 277.

\bibitem{Fisk2}       Z. Fisk and G. W. Webb,
			Phys. Rev. Lett. {\bf 36} (1976) 1084.

\bibitem{Gurvitch}	M. Gurvitch, A. K. Ghosh, B. L. Gyorffy, H. Lutz,
			O. F. Kammerer, J. S. Rosner, and M. Strongin,
			Phys. Rev. Lett. {\bf 41} (1978) 1616.

\bibitem{Wiesmann}	H. Wiesmann, M. Gurvitch, H. Lutz, B. Schwartz,
        M. Strongin, P.B. Allen and J.W. Halley,
        Phys. Rev. Lett. {\bf 38} (1977) 782.

\bibitem{Gurvitch2}	M. Gurvitch,
			Phys. Rev. B {\bf 24} (1981) 7404.

\bibitem{Allen}		P.B. Allen,
        in {\sl Superconductivity in d- and f-Band Metals},
        H. Suhl and M.B. Maple, eds.
        (Academic Press, NY 1980) p. 291;
			P.B. Allen,
        in {\sl Physics of Transition Metals, Leeds, 1980}
       (Inst. Phys. Conf. Ser. No. 55, 1981) p. 425.


\bibitem{Nikolic}	B. Nikoli\'c and P. B. Allen,
			Phys. Rev. B {\bf 63} (2000) 020201.

\bibitem{Slevin}      K. Slevin, T. Ohtsuki, and T. Kawarabayashi, 
			Phys. Rev. Lett. {\bf 84} (2000) 3915.

\bibitem{Gurvitch3}	M. Gurvitch and A. T. Fiory,
			Phys. Rev. Lett. {\bf 59} (1987) 1337.

\bibitem{Emery}		V. J. Emery and S. A. Kivelson,
			Phys. Rev. Lett. {\bf 74} (1995) 3253.

\bibitem{Hou}		J. G. Hou, L. Lu, V. H. Crespi, X.-D. Xiang,
			A. Zettl, and M. L. Cohen,
			Solid State Commun. {\bf 93} (1995) 973.

\bibitem{Pavuna}	D. Pavuna, H. Berger, and L. Forro,
			J. Eur. Ceram. Soc {\bf 19} (1999) 1518.

\bibitem{Tyler}		A. W. Tyler, A. P. Mackenzie, S. NishiZaki,
			and Y. Maeno,
			Phys. Rev. B {\bf 58} (1998) 10107.

\bibitem{Khalifah}	P. Khalifah, K. D. Nelson, R. Jin, Z. Q. Mao, Y. Liu,
			Q. Huang, X. P. A. Gao, A. P. Ramirez, and R. J. Cava,
			Nature {\bf 411} (2001) 669.

\bibitem{Kiv}		S. A. Kivelson, E. Fradkin, and V. J. Emery,
			Nature {\bf 393} (1998) 550.

\bibitem{Arnason}	S. B. Arnason, S. P. Herschfield, and A. F. Hebard,
			Phys. Rev. Lett. {\bf 81} (1998) 3936.

\bibitem{Millis}	A. J. Millis, L. Hu, and S. Das Sarma,
			Phys. Rev. Lett. {\bf 82} (1999) 2354.

\bibitem{Merino}	J. Merino, and R. H. McKenzie,
			Phys. Rev. B {\bf 61} (2000) 7996.

\bibitem{Gunnarsson1}	O. Gunnarsson and J. E. Han,
			Nature {\bf 405} (2000) 1027.

\bibitem{Gunnarsson2}	M. Calandra and O. Gunnarsson,
			cond-mat/0106397.

\bibitem{Chakraborty}	B. Chakraborty and P.B. Allen,
        		Phys. Rev. Lett. {\bf 42} (1979) 736;
			P.B. Allen and B. Chakraborty,
        		Phys. Rev. B {\bf 23} (1981) 4815.

\bibitem{Ando}		Y. Ando, A. N. Lavrov, S. Komiya,
			K. Segawa, and X. F. Sun,
			Phys. Rev. Lett. {\bf 87} (2001) 017001.


\end{references}
\end{document}